\documentclass[12pt]{article}

\usepackage{epsf}
\usepackage[dvips]{graphicx}

\begin{document}
\begin{center}
{\bfseries DEUTERON-PROTON CHARGE EXCHANGE REACTION AT SMALL TRANSFER
MOMENTUM}

\vskip 5mm

N.B. Ladygina$^{1 \dag}$, A.V.Shebeko$^{2}$

\vskip 5mm
{\small
(1) {\it
Laboratory of High Energies,
Joint Institute for Nuclear Research, 141980 Dubna, Russia,
}
\\
(2) {\it
NSC Kharkov Institute of Physics \& Technology, 61108 Kharkov, Ukraine
}
\\


$\dag$ {\it
E-mail: ladygina@sunhe.jinr.ru
}}
\end{center}

\vskip 5mm

\begin{center}
\begin{minipage}{150mm}
\centerline{\bf Abstract}
The charge-exchange reaction $pd\to npp$ at 1 GeV projectile proton
energy is studied. This reaction is considered in a special kinematics,
when the transfer momentum from the beam proton to fast outgoing neutron
is close to zero. 
Our approach is based on the Alt-Grassberger-Sandhas
formulation of the multiple-scattering theory for the three-nucleon system.
The matrix inversion method has been applied to take account of the
final state interaction (FSI) contributions. The differential cross section,
tensor analyzing power $C_{0,yy}$, vector-vector $C_{y,y}$ and
vector-tensor $C_{y,xz}$ spin correlation parameters of the initial
particles are presented. It is shown, that the FSI effects play a very
important role under such kinematical conditions. The high sensitivity
of the considered observables to the elementary nucleon-nucleon
amplitudes has been obtained.
\end{minipage}
\end{center}

\vskip 10mm

\section{Introduction}

During the  last decades the deuteron- proton charge exchange
reaction has been studied both from the experimental and theoretical
point of view. A considerable interest in this reaction is connected,
first of all, to the opportunity to extract some information about the
spin-dependent part of the elementary nucleon-nucleon charge
exchange amplitudes. This idea was suggested by Pomeranchuk \cite {Pom}
already in 1951, but until now it continues to be of interest. 
Later, this supposition
has been developed in \cite {Dean, Wil, Car}. It was shown, that in the
plane-wave impulse approximation (PWIA) the differential cross
section and tensor analyzing power $T_{20}$ in the dp-charge exchange
reaction are actually fully determined  by the spin-dependent
part of the elementary $np\to pn$
 amplitudes.

Nowadays the experiment on the study of the dp-charge exchange reaction at
the small transfer momentum in the GeV-region is planned at ANKE setup
 at COSY \cite {cosy}. The aim of this experiment is to provide 
information about  spin-dependent np-elastic scattering
amplitudes in the energy
region  where  phase-shift analysis data are absent. 

From our point
of view, under kinematical conditions proposed in this experiment,
when momentum of the emitted neutron has the same direction and magnitude
as the beam proton (in the deuteron rest frame), and relative momentum
of  two protons is very small, the final state interaction (FSI)
effects play very important role. The contribution of the D-wave
in the DWF into differential cross section in this kinematics
must be negligible \cite {vvg}. However, for the polarization observables
the influence of the D-component can be significant.

The goal of our paper is to study the importance of the D-wave and
FSI effects under kinematical conditions of the planned experiment.
We consider $pd\to npp$ reaction in the approach, which has been used
by us to describe the pd breakup process at 1 GeV projectile proton
energy \cite {LSh}. This approach is based on the Alt-Grassberger-Sandhas
 formulation
of  the multiple-scattering theory for the three-nucleon system.
The matrix inversion method 
has been applied to take account of the FSI contributions. Since
unpolarized and polarized mode of the deuteron beam are supposed to be
employed 
in the experiment, we also calculate both the differential cross
section and a set of the polarization observables.
It should be noted, in this paper we have not considered the Coulomb
interaction in the (pp)-pair. This problem is nontrivial and
requires a special investigation.

\section{Theoretial formalism}

In accordance to the three-body collision theory, the amplitude of 
the deuteron proton charge exchange reaction,
\begin{eqnarray}
p(\vec p)+d(\vec 0)\to n(\vec p_1)+p(\vec p_2)+p(\vec p_3)
\end{eqnarray}
is defined by the matrix element of the transition operator $U_{01}$
\begin{eqnarray}
\label{ampl}
{U}_{pd\to ppn} \equiv \sqrt {2} <123|[1-(1,2)-(1,3)] U_{01}|1(23)>=
\delta (\vec p -\vec p_1-\vec p_2 -\vec p_3){\cal J}.
\end{eqnarray}
As consequence  of the particle identity in initial and final states
 the permutation operators for two nucleons $(i,j)$ appear in this expression.

As was shown in   ref.\cite {LSh} the matrix element 
$U_{pd \to npp}$ can be presented as

\begin{eqnarray}
\label{am}
U_{pd \to npp}&=&\sqrt {2} <123|[1-(2,3)][1+t_{23}(E-E_1) 
g_{23} (E-E_1)]t_{12}^{sym}|1(23)>,
\end{eqnarray}
where the operator $g_{23} (E-E_1)$ is a free propagator for the
(23)-subsystem and the scattering operator $t_{23}(E-E_1)$
satisfies the Lippmann-Schwinger (LS)
equation with two-body force operator $V_{23}$ as  driving term
\begin{eqnarray}
\label{LS}
t_{23}(E-E_1) = V_{23} + V_{23} g_{23}(E-E_1) t_{23}(E-E_1) .
\end{eqnarray}
Here $E$ is the total energy of the three-nucleon system 
$E=E_1+E_2+E_3$.

Let us rewrite the matrix element (\ref{am}) indicating explicitly
the particle quantum numbers,
\begin{eqnarray}
U_{pd\to npp}=\sqrt {2}
<\vec {p_1} m_1 \tau_1,\vec {p_2} m_2 \tau_2,\vec {p_3} m_3 \tau_3|
[1-(2,3)] \omega_{23} t^{sym}_{12} |\vec {p} m \tau ,\psi _{1 M_d 0 0} (23)>,
\nonumber
\end{eqnarray}
where $\omega_{23}=[1+t_{23}(E-E_1) g_{23} (E-E_1)]$ and
the the spin and isospin  projections denoted as
$m$ and $\tau$, respectively.  The operator $t_{12}^{sym}$ is symmetrized
NN-operator, $t_{12}^{sym}=[1-(1,2)]t_{12}$.

In this paper we consider the special kinematics, when
transfer momentum  $\vec q=\vec p -\vec p_1 $ is close to zero.
In other words, the neutron momentum has the same value and
direction as the beam proton. In fact, since the difference between
proton and neutron masses and deuteron binding energy take place,
the transfer momentum is not  exactly zero, $q\approx 1.8$ MeV/c. But 
because of this value is very small and has no significant influence
on the results, we shall suppose $q=0$ in the subsequent calculations.

Under such kinematical conditions one can anticipate
that the FSI in the $^1S_0$ state is prevalent at comparatively small
$p_0$-values.
In such a way we get the following expression for amplitude of
the dp charge exchange process \cite {EPJ}
\begin{eqnarray}
\label{ampl}
{\cal J}&=&{\cal J}_{PWIA}+{\cal J }_{^1S_0}
\nonumber\\
\nonumber\\
{\cal J}_{PWIA}&=& <L M_L 1 {\cal M_S}|1 M_D>
u_L ( p_0 )
Y_L^{M_L}(\widehat { p_0})
\nonumber\\
&&\Bigl\{ <{1\over 2} m_2^\prime {1\over 2} m_3|1 {\cal M_S}>
< m_1 m_2|
t^0 (\vec p,\vec p_0) -t^1(\vec p,\vec p_0)
| m m_2 ^ \prime >-
\nonumber\\
&&<{1\over 2} m_2^\prime {1\over 2} m_2|1 {\cal M_S}>
< m_1 m_3|
t^0(\vec p,\vec p_0) -t^1(\vec p,\vec p_0)
| m m_2 ^ \prime > \}
\\
\nonumber\\
{\cal J}_{^1S_0}&=&\frac {(-1)^{1-m_2 -m_2^\prime}}{\sqrt {4\pi }} 
\delta _{m_2 ~ -m_3}
<{1\over 2} m^{\prime\prime } {1\over 2} -m_2^\prime|1 M_D> 
\nonumber\\
&&< m_1 m _2^\prime |
t^0(\vec p\vec p_0^\prime) -t^1(\vec p,\vec p_0^\prime)
| m m ^ {\prime\prime } >
\int dp _0 {^\prime } p _0 {^\prime }	^2
\psi _{00} ^{001} (p_0^\prime ) u_0(p_0^\prime).
\end{eqnarray}
The wave function of the final
$pp$-pair  $\psi _{00} ^{001} (p_0^\prime )$ can be expressed by a series of
$\delta$-functions, what enables us to perform the integration
over $p_0^\prime$ in this expression.
We use the phenomenological model suggested by Love and Franey \cite {LF}
for
description the high energy nucleon-nucleon matrix $t(\vec p,\vec p_0^\prime)$.

\section{Results and discussions}

We define general spin observable related with polarization of 
initial particles in terms of the Pauli $2 \times 2$ spin matrices $\sigma$
for the proton and a set of spin operators $S$ for deuteron 
as following
\begin{eqnarray}
C_{\alpha\beta}=\frac {Tr ({\cal J}\sigma _\alpha S_\beta {\cal J})}
{Tr ({\cal J} {\cal J}^+) },
\end{eqnarray}
where indices $\alpha$ and $\beta$ refer to the proton and deuteron
polarization, respectively; $\sigma _0$ and $S_0$ corresponding to the
unpolarized particles are the unit matrices of two and three dimensions.
In such a way, Eqs.(\ref {ampl}) for dp- charge exchange
amplitude  enables us  to get the relation for any variable of this
process taking into account two slow protons final state interaction
in $^1S_0$ -state. So, we have following expression for the spin- averaged
squared amplitude in kinematics, when
one of the slow protons is emitted along the beam direction as well as
neutron $(\theta_2=0^0)$ 
\begin{eqnarray}
\label{c0}
C_0&=&{1\over {2 \pi}}\left( \frac{m_N+E_p}{2E_p}
\right) ^2\{ (2B^2+F^2)({\cal U}^2(p_2)+w^2(p_2))+
\\
&&(F^2-B^2)w(p_2)(w(p_2)-2\sqrt 2 Re{\cal U}(p_2))\},
\nonumber
\end{eqnarray}
where ${\cal U}(p_2)=u(p_2)+\int dp_0^\prime {p_0^\prime }^2
\psi _{00}^{001}(p_0^\prime ) u(p_0^\prime )$
is the S-component of the DWF $u(p_2)$  corrected on the FSI of 
the (pp)-pair and $w(p_2)$ is the D-component of the DWF;
$B$ and $F$ are the spin dependent nucleon-nucleon amplitudes \cite {LF}.

We use a right-hand coordinate system defined in accordance to the 
Madison convention \cite {mad}. The quantization $z$-axis is along
the beam proton momentum $\vec p$. Since the direction of 
$\vec p \times \vec p_1$ is undefined in the collinear geometry, 
we choose the $y$-axis normal to  the beam momentum. Then
third axis is $\vec x =\vec y\times \vec z$.

The tensor analyzing power can be presented in the following form 
\begin{eqnarray}
C_{0,yy}\cdot C_0&=&{1\over {4 \pi}}\left( \frac{m_N+E_p}{2E_p}
\right) ^2
\{2(F^2-B^2)({\cal U}^2(p_2)+w^2(p_2))+
\\
&&(2F^2+B^2)w(p_2)(w(p_2)-2\sqrt 2 Re {\cal U}(p_2))\}
\nonumber
\end{eqnarray}
Note, that only squared nucleon- nucleon spin- flip amplitudes
$B^2$ and $F^2$ are in expression for the tensor analyzing power $C_{0,yy}$
and differential cross section. However, the spin correlation
due to vector polarization of deuteron and beam proton
 contains the interference
terms of this amplitudes
\begin{eqnarray}
C_{y,y}\cdot C_0&=&-{2 \over {4 \pi}}\left( \frac{m_N+E_p}{2E_p}
\right) ^2
\{Re(FB^*)[2{\cal U}^2(p_2)-2w^2(p_2)-
\sqrt 2 Re{\cal U}(p_2)w(p_2)]-
\nonumber\\
&&3\sqrt 2 Im(FB^*)Im{\cal U}(p_2)w(p_2)\}
\end{eqnarray} 
It is interesting, that there is the term proportional to the imaginary part
of ${\cal U}(p_2)$. It has a non-zero value only in case when FSI
is taken into account. The analogous result we have obtained for the
vector-tensor spin correlation
\begin{eqnarray}
\label{cyxz}
C_{y,xz}\cdot C_0&=&-{3 \over {4 \pi}}\left( \frac{m_N+E_p}{2E_p}
\right) ^2
\{Im(FB^*)[2{\cal U}^2(p_2)-2w^2(p_2)-
\sqrt 2 Re{\cal U}(p_2)w(p_2)]+
\nonumber\\
&&3\sqrt 2
Re(FB^*)Im{\cal U}(p_2)w(p_2)\}
\end{eqnarray}

The differential cross section and three polarization
observables are presented in  figs.(1-4). 
The Love and Franey parametrization with a set of parameters
obtained by fitting of the modern phase shift data SP00 \cite {ar, said}
has been employed for NN-amplitude.
 All calculations were carried out with
Paris NN-potential \cite {NN} and Paris DWF \cite {Par}.

One can see, the FSI contribution to the differential cross section (fig.1)
is significant even at the very small proton momentum, while for the 
polarization observables the difference between PWIA and PWIA+FSI
is visible only for $p_2 \ge 10-15 $ MeV/c.  However, with increase
of the proton momentum up to 50 MeV/c the importance of the FSI
corrections to the PWIA also increases.

Note, the absolute value of the tensor analyzing power $C_{0,yy}$ (fig.2)
in the momentum interval of interest is near zero. In order to
understand the source of that, we disregard the D-wave in the DWF.
Then  the polarization
observables are defined
by the ratio of the nucleon-nucleon charge exchange amplitudes only
\begin{eqnarray}
\label{w0}
C_{0,yy}&=&{1\over 2}\cdot \frac {F^2-B^2}{2B^2+F^2}
\nonumber\\
C_{y,y}&=&-2\cdot \frac {Re(FB^*)}{2B^2+F^2}
\\
C_{y,xz}&=&-3\cdot\frac {Im(FB^*)}{2B^2+F^2}
\nonumber
\end{eqnarray}
Thus, the nearness of the tensor analyzing power to zero  indicates that
the absolute values of the spin-flip NN-amplitudes approximately
equal each other, $|B|\approx |F|$.

The vector-tensor spin correlation
$C_{y,xz}$ (fig.4) has also very small value,
$|C_{y,xz}|\approx 0.06$.
The magnitude of this observable decreases
up to zero for $p_2\approx 50$ MeV/c, if the FSI corrections and
D-wave in the deuteron are taken into account, while it is almost
 constant in the PWIA and PWIA+FSI without D-wave.
One can see from Eqs.(\ref {cyxz}, \ref {w0}) for $C_{y,xz}$, the
reason of this behaviour  is connected with the small
value of the imaginary part of the nucleon-nucleon amplitudes
product, $Im (FB^*)$. In such a way, the great contribution
into $C_{y,xz}$ gives the term proportional to $Re (FB^*)$,
which defined by D-wave and imaginary part of the generalized
function ${\cal U}(p_2)$. Note, that $Im {\cal U}(p_2)\ne 0$, if
FSI taken into account.

The other situation is for the vector-vector spin correlation 
$C_{y,y}$ (fig.3). The term
proportional to $Re(FB^*)$ gives also
a considerable contribution in this observable , but it is multiplied
on the ${\cal U}^2(p_2)$.
The magnitude of $C_{y,y}$ is close to the theoretical limit -2/3,
that confirms to the
conclusion about approximate equality of the nucleon-nucleon
amplitudes, $|B|$ and $|F|$. Besides, this allows to conclude, that the relative
phase between these amplitudes is close to zero.
It is seen from Eq.(\ref {w0}), where D-wave
was neglected.


\begin{figure}[t]
\begin{minipage}{7.5cm}
 \epsfysize=90mm
 \epsfbox{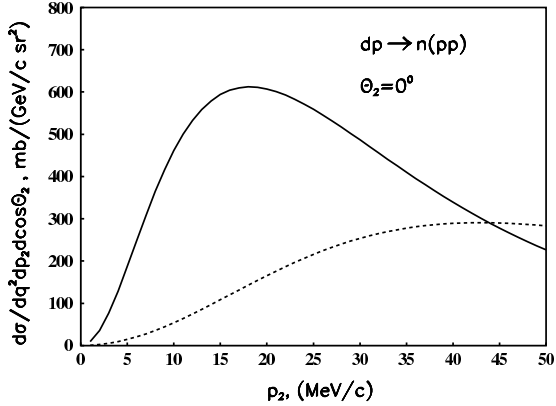}
 \vspace*{-3cm}
 \caption{
 The differential cross section at $\vec q=0$ as 
a function of one of the slow proton momentum. The dashed and full line correspond
to the PWIA and PWIA+FSI, respectively.
}
 \end{minipage}
\end{figure}

\begin{figure}[t]
\vspace*{-9.9cm}
\hfill {
\begin{minipage}{7.5cm}
 \epsfysize=90mm
 \epsfbox{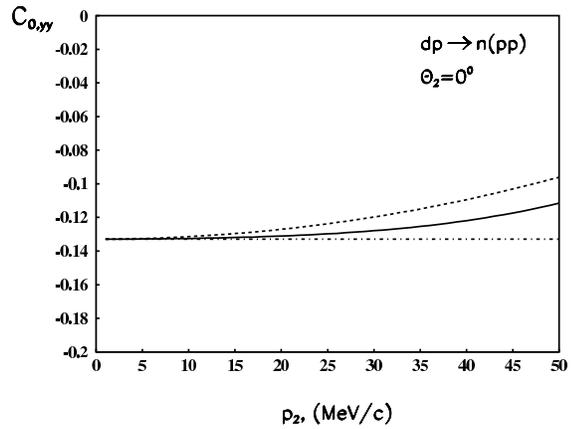}
 \vspace*{-3cm}
 \caption{
 The tensor analyzing power $C_{yy}$ vs. $p_2$. The dashed line corresponds 
to  PWIA; dash-dotted and full lines are PWIA+FSI without D-component in the DWF
and with it, respectively.
}
\end{minipage}
}
\end{figure}

\begin{figure}[t]
\begin{minipage}{7.5cm}
 \epsfysize=90mm
 \epsfbox{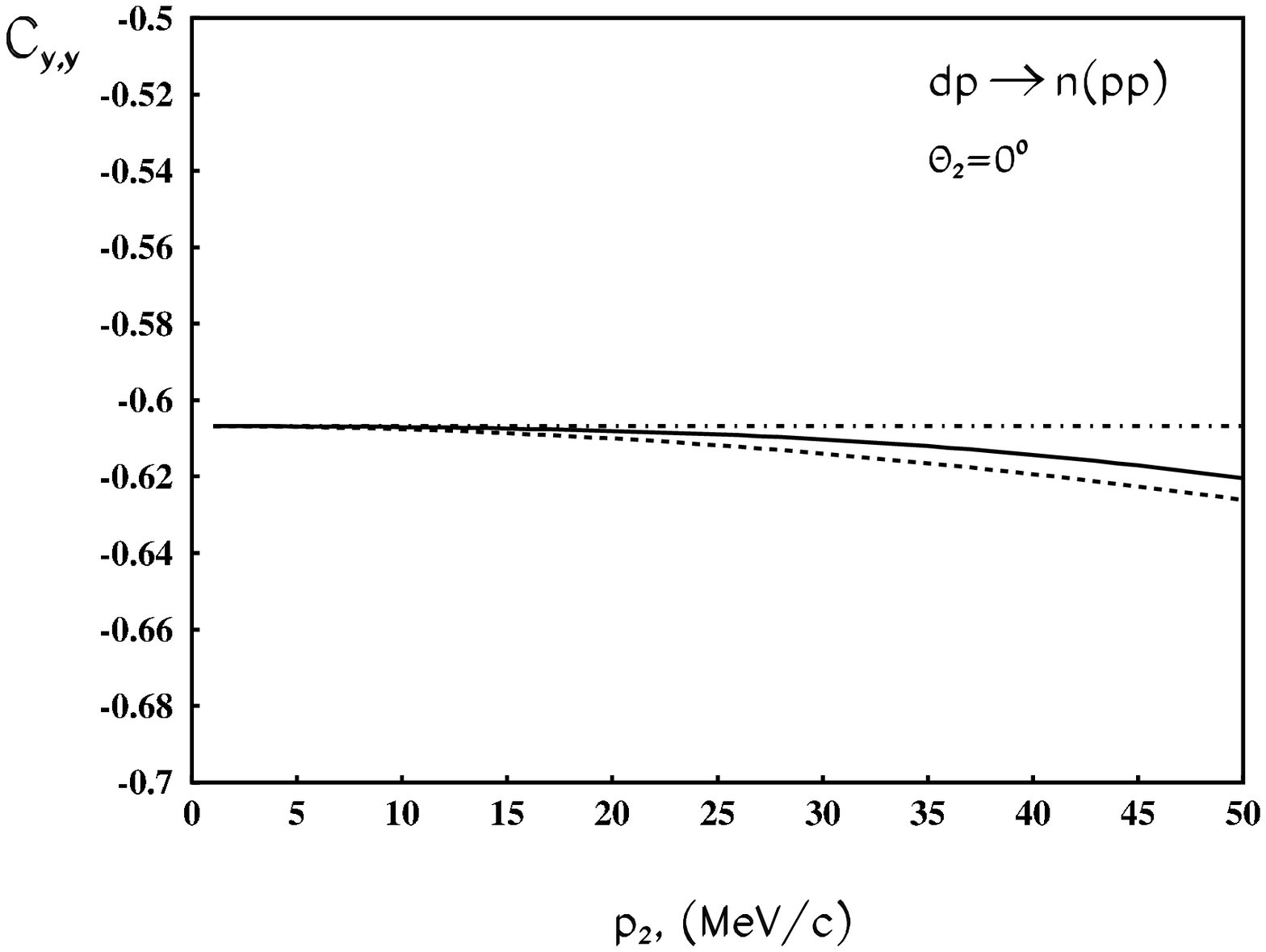}
 \vskip -3cm
 \caption{
 The spin-correlation $C_{y,y}$ due to the vector polarization of 
the deuteron. The curves are the same as in fig.2.
}
 \end{minipage}
\end{figure}

\begin{figure}[t]
\vspace*{-8.4cm}
\hfill {
\begin{minipage}{7.5cm}
 \epsfysize=90mm
 \epsfbox{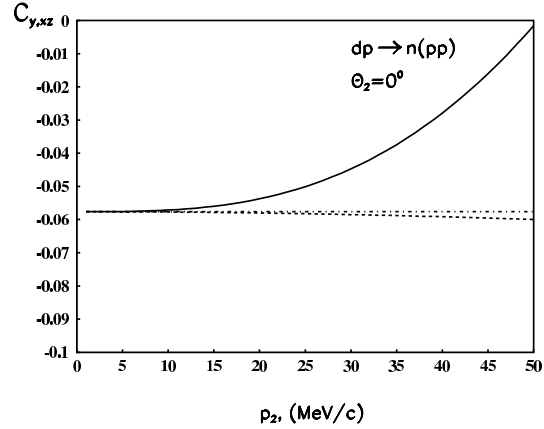}
 \vskip -3cm
 \caption{
 The spin-correlation $C_{y,xz}$ due to the tensor polarization of
the deuteron. The curves are the same as in fig.2.
}
\end{minipage}}
\end{figure}

\section{Conclusion}

We have studied the deuteron -proton charge exchange reaction at
1 GeV energy in special kinematics, $\vec q \approx 0$. The influence
of the  D-wave in the deuteron and
FSI between two slow protons has been considered. 
It was shown, that D-wave and FSI
effects  are negligible for the polarization observables at 
proton momentum up to 10-15 MeV/c. As a result, in this region the
polarization observables are defined by the ratio of the nucleon-
nucleon charge exchange amplitudes only.
However, it  should not be ignored that the
 importance of the D-wave and ,
especially, FSI into polarization observables increases at 
$p_2 \ge 15$ MeV/c. In such a way, we conclude, that the
 ratio of the nucleon-
nucleon charge exchange amplitudes and phase shift between them can
be extracted from experimental data rather simple, if the experimental
conditions and technical setup possibilities
allow to work in this small momentum interval. In the opposite case,
this procedure is more complicated and  model dependent.
 It should be remembered 
that the FSI contribution to the differential cross section is very
significant in comparison with PWIA predictions even at very small
proton momentum. This fact does not enable us to get the absolute
value of the nucleon-nucleon spin flip amplitudes without considering
the FSI corrections.

\vspace{2cm}
We are grateful to V.V.Glagolev, M.S.Nioradze and A.Kacharava for
inspiration of interest to this problem.
The authors are thankful to V.P. Ladygin for fruitful discussions. 

\end{document}